\pdfoutput=1
\documentclass[sigconf,x11names]{acmart}

\newcommand{\ca}[1]{{\color{black}{#1}}}

\usepackage{xcolor}

\usepackage{tikz}
\usepackage{pgfplots}
\usepackage{animate}
\usetikzlibrary{calc}
\usetikzlibrary{positioning}
\usetikzlibrary{shapes,arrows, positioning, calc}
\usetikzlibrary{overlay-beamer-styles}
\usetikzlibrary{chains,shapes.multipart}
\usetikzlibrary{scopes}
\usetikzlibrary{automata}
\usetikzlibrary{positioning}  %                 ...positioning nodes
\usetikzlibrary{arrows}       %                 ...customizing arrows
\usetikzlibrary{intersections}

\usepackage{glossaries}

\newacronym{aoa}{AOA}{Angle Of Arrival}
\newacronym{ac}{AC}{Access Category}
\newacronym{ack}{ACK}{Acknowledgement}
\newacronym{akmp}{AKMP}{Authentication and Key Management Protocol}
\newacronym{a-mpdu}{A-MPDU}{Aggregated MPDU}
\newacronym{aod}{AOD}{Angle Of Departure}
\newacronym{ap}{AP}{Access Point}
\newacronym{asap}{ASAP}{As Soon As Possible}
\newacronym{awgn}{AWGN}{Additive White Gaussian Noise}
\newacronym{awv}{AWV}{Antenna Weight Vector}
\newacronym{brp}{BRP}{Beam Refinement Phase}
\newacronym{bsr}{BSR}{Buffer Status Report}
\newacronym{bss}{BSS}{Basic Service Set}
\newacronym{cdf}{CDF}{Cumulative distribution function}
\newacronym{cli}{CLI}{Command Line Interface}
\newacronym{cots}{COTS}{Commercial off-the-shelf}
\newacronym{csi}{CSI}{Channel State Information}
\newacronym{dhss}{DHss}{Diffie-Hellman (DH) shared secret}
\newacronym{dl}{DL}{Downlink}
\newacronym{dmg}{DMG}{Directional Multi-Gigabit}
\newacronym{edca}{EDCA}{Enhanced Distributed Channel Access}
\newacronym{edmg}{EDMG}{Enhanced Directional Multi-Gigabit}
\newacronym{fft}{FFT}{Fast Fourier Transform}
\newacronym{fpbt}{FPBT}{First Path Beamforming Training}
\newacronym{ftm}{FTM}{Fine Timing Measurement}
\newacronym{ftmr}{FTMR}{FTM Request}
\newacronym{gi}{GI}{Guard Interval}
\newacronym{gcl}{GCL}{Gate Control List}
\newacronym{hkdf}{HKDF}{HMAC-based Key Derivation Function}
\newacronym{i2r}{I2R}{Initiator-to-Responder}
\newacronym{ista}{ISTA}{Initiating Station}
\newacronym{iftm}{IFTM}{Initial FTM}
\newacronym{iftmr}{IFTMR}{Initial FTM Request}
\newacronym{ipftmr}{IPFTMR}{Initial Protected FTM Request}
\newacronym{kdf}{KDF}{Key Derivation Function}
\newacronym{lci}{LCI}{Location Configuration Information}
\newacronym{ldpc}{LDPC}{Low Density Parity Check}
\newacronym{los}{LOS}{Line Of Sight}
\newacronym{mcs}{MCS}{Modulation Coding Scheme}
\newacronym{mic}{MIC}{Message Integrity Code}
\newacronym{ml}{ML}{Machine Learning}
\newacronym{mmwave}{mmWave}{Millimiter Wave}
\newacronym{mpdu}{MPDU}{Medium Access Control (MAC) protocol data unit}
\newacronym{mu}{MU}{Multi User transmission}
\newacronym{ngp}{NGP}{Next Generation Positioning}
\newacronym{nlos}{NLOS}{Non Line Of Sight}
\newacronym{ntb}{NTB}{Non-Triggered-Based}
\newacronym{obss}{OBSS}{Overlapping BSS}
\newacronym{ofdm}{OFDM}{Orthogonal Frequency Division Multiplexing}
\newacronym{ofdma}{OFDMA}{Orthogonal Frequency Division Multiple Access}
\newacronym{p2p}{P2P}{Peer to Peer}
\newacronym{pasn}{PASN}{Pre-Association Security Negotiation}
\newacronym{pdp}{PDP}{Power Delay Profile}
\newacronym{per}{PER}{Packet Error Rate}
\newacronym{pftm}{PFTM}{Protected FTM}
\newacronym{phy}{PHY}{physical}
\newacronym{pmk}{PMK}{Pairwise Master Key}
\newacronym{pmksa}{PMKSA}{Pairwise Master Key Security Association}
\newacronym{ppdu}{PPDU}{Physical Protocol Data Unit}
\newacronym{prf}{PRF}{Pseudo-random Function}
\newacronym{prk}{PRK}{Pseudo-random Key}
\newacronym{ptb}{PTB}{Passive Triggered-Based}
\newacronym{ptk}{PTK}{Pairwise Transient Key}
\newacronym{ptksa}{PTKSA}{Pairwise Transient Key Security Association}
\newacronym{qos}{QoS}{Quality of Service}
\newacronym{r2i}{R2I}{Responder-to-Initiator}
\newacronym{rbir}{RBIR}{Received Bit mutual Information Rate}
\newacronym{rsna}{RSNA}{Robust Security Network Association}
\newacronym{rsta}{RSTA}{Responding Station}
\newacronym{rssi}{RSSI}{Received Signal Strength Indicator}
\newacronym{rtt}{RTT}{Round Trip Time}
\newacronym{rtwt}{rTWT}{restricted Target Wake Time}
\newacronym{ru}{RU}{Resource Unit}
\newacronym{siso}{SISO}{Single Input Single Output}
\newacronym{sta}{STA}{Station}
\newacronym{sinr}{SINR}{Signal to Interference and Noise Ratio}
\newacronym{snr}{SNR}{Signal Noise Ratio}
\newacronym{scofdm}{SC-OFDM}{Single Carrier Orthogonal Frequency-Division Multiplexing}
\newacronym{tas}{TAS}{Time-Aware Shaper}
\newacronym{tb}{TB}{Triggered-Based}
\newacronym{tdma}{TDMA}{Time Division Multiple Access}
\newacronym{tgax}{TGax}{802.11ax Task Group}
\newacronym{toa}{TOA}{Time of Arrival}
\newacronym{tod}{TOD}{Time of Departure}
\newacronym{tof}{ToF}{Time of Flight}
\newacronym{trn}{TRN}{Training}
\newacronym{tsf}{TSF}{Time Synchronization Function}
\newacronym{tsn}{TSN}{Time-Sensitive Networking}
\newacronym{txss}{TXSS}{Transmit sector sweep}
\newacronym{twt}{TWT}{Target Wake Time}
\newacronym{ul}{UL}{Uplink}

\newtheorem{theorem}{Theorem}
\newtheorem{corollary}{Corollary}[theorem]
\AtBeginDocument{%
  }

\setcopyright{none}
\copyrightyear{2023}
\acmYear{2023}
\acmDOI{}

\acmConference{}{}{}
\acmPrice{}
\acmISBN{}

\pgfplotsset{compat=1.18}

\begin{document}

%\title{To TWT, or not to TWT,
%that is the question}
\title{Aligning rTWT with 802.1Qbv: a Network Calculus Approach}

\author{Carlos Barroso-Fernández}
\email{cbarroso@pa.uc3m.es}
\affiliation{%
  \institution{Universidad Carlos III de Madrid}
  \streetaddress{Avda. Universidad, 30}
  \city{Leganés}
  \state{Madrid}
  \country{Spain}
  \postcode{28911}
}
\author{Jorge Martín-Pérez}
\orcid{0000-0001-9295-1601}
\email{jorge.martin.perez@upm.es}
\affiliation{%
  \institution{Universidad Politécnica de Madrid}
  \streetaddress{Avda. Complutense, 30}
  \city{Madrid}
  \state{Madrid}
  \country{Spain}
  \postcode{28040}
}
\author{Constantine Ayimba}
\email{ayconsta@it.uc3m.es}
\affiliation{%
  \institution{Universidad Carlos III de Madrid}
  \streetaddress{Avda. Universidad, 30}
  \city{Leganés}
  \state{Madrid}
  \country{Spain}
  \postcode{28911}
}
\author{Antonio de la Oliva}
\email{aoliva@it.uc3m.es}
\affiliation{%
  \institution{Universidad Carlos III de Madrid}
  \streetaddress{Avda. Universidad, 30}
  \city{Leganés}
  \state{Madrid}
  \country{Spain}
  \postcode{28911}
}
\renewcommand{\shortauthors}{Barroso et al.}

\begin{abstract}
    \ca{Industry 4.0 applications impose the challenging demand of delivering packets with bounded latencies via a wireless network. This is further complicated if the network is not dedicated to the time critical application. In this paper we use network calculus analysis to derive closed form expressions of latency bounds for time critical traffic when 802.11 \gls{twt} and 802.1Qbv work together in a shared 802.11 network.}
\end{abstract}





\maketitle

\section{Introduction}

\ca{Unscheduled channel access in 802.11 networks negatively impacts their reliability particularly in dense deployments. This is a particularly challenging phenomenon given the imminent use of wireless networks for industry 4.0 applications~\cite{industry4_0}. %Recent improvements proposed in 802.11ax regarding 
\gls{twt} reduces contention by allowing the AP to instruct \gls{sta}s to turn their transceivers on or off at agreed intervals. The discipline thus transforms a dense deployment to one with a manageable number of \gls{sta}s during a \gls{twt} session therefore reducing contention and improving the radio channel access delay. 
Moreover, 802.11be will extend it to \gls{rtwt} to completely eliminate contention\footnote{In 802.11ax contention can be ameliorated with techniques such as trigger-based \gls{ofdma}.}. %Maybe we need to change rTWT to TWT.
%The standard allows \gls{ofdma} in both uplink~(UL) and downlink~(DL) therefore \gls{sta}s can be assigned different \gls{ru}s for transmission in both directions. 
This method has been proposed as a scheduling mechanism in \cite{Chen_TWTscheduler} that achieves improvement in overall network throughput but does not consider the priority of packets. This approach is therefore inadequate for \gls{tsn} flows requiring bounded latency. We propose an enhancement, incorporating \gls{tsn}, by aligning channel access via \gls{rtwt} with 802.1Qbv scheduling. The authors of~\cite{TWT_PoC} provide a proof of concept implementation of \gls{twt} for time aware scheduling that isolates priority traffic from best effort traffic achieving promising results on bounded latency and jitter. We leverage network calculus analysis~\cite{leboudec} to prove that bounded latencies, required by \gls{tsn}, can be achieved using this approach when both high and low priority packets are present in a \gls{sta} with some violation probability that depends on channel quality. In Section~\ref{sec:Qbv_tdma} we model the 802.1Qbv opening gates using Network Calculus~\cite{netcal8021qbv,leboudec}, and in Section~\ref{subsec:concurrent} the wireless errors using stochastic scaling~\cite{stochastic-scaling}. We discuss results in Section~\ref{sec:results} and conclusions in Section~\ref{sec:conclusions}.}

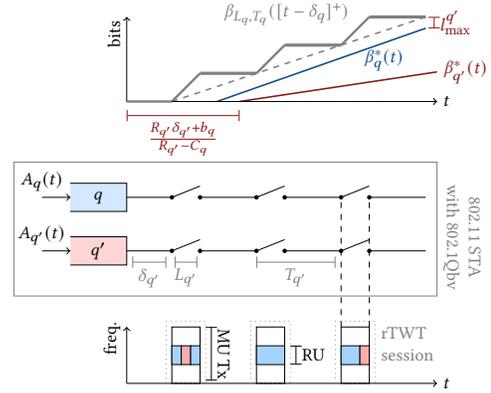
\begin{figure}[t]
    \scalebox{0.75}{\begin{tikzpicture}

    % queue
    \pgfmathsetmacro\qy{-1.7}
    \pgfmathsetmacro\qx{-.5}
    \pgfmathsetmacro\qw{.5} % 1/2 queue width
    \pgfmathsetmacro\qh{.25} % 1/2 queue height
    \coordinate (queue) at (\qx,\qy);
    \path[fill=DodgerBlue1!20,draw] ($(queue)+(-\qw,\qh)$) --
        ($(queue)+(\qw, \qh)$) --
        ($(queue)+(\qw,-\qh)$) --
        ($(queue)+(-\qw,-\qh)$);
    \node at (queue) {$q$};

    % Arrival flow
    \draw[->] ($(queue)-(2*\qw,0)$)
        node[anchor=south] {$A_q(t)$}
        -- ($(queue)-(\qw,0)$);

    % low priority queue q2
    \pgfmathsetmacro\qly{-2.65}
    \pgfmathsetmacro\qlx{-.5}
    \pgfmathsetmacro\qlw{.5} % 1/2 queue width
    \pgfmathsetmacro\qlh{.25} % 1/2 queue height
    \pgfmathsetmacro\llmax{.2} % max low packet
    \coordinate (queuel) at (\qlx,\qly);
    \path[fill=Firebrick1!20,draw] ($(queuel)+(-\qw,\qh)$) --
        ($(queuel)+(\qlw, \qlh)$) --
        ($(queuel)+(\qlw,-\qlh)$) --
        ($(queuel)+(-\qlw,-\qlh)$);
    \node at (queuel) {$q'$};

    % Arrival flow low priority
    \draw[->] ($(queuel)-(2*\qlw,0)$)
        node[anchor=south] {$A_{q'}(t)$}
        -- ($(queuel)-(\qlw,0)$);

    % Opening gates params
    \pgfmathsetmacro\TmL{1} % T-L
    \pgfmathsetmacro\L{.5} % L
    \pgfmathsetmacro\h{.5} % TDMA height
    \pgfmathsetmacro\deltaA{.8} % deltaA
    \pgfmathsetmacro\steps{2}

    % Slope params high prio
    \pgfmathsetmacro\C{\h/\L} % channel rate
    \pgfmathsetmacro\Cq{.1*\C} % q arriving rate
    \pgfmathsetmacro\bq{.2*\h} % q burst
    \pgfmathsetmacro\Rq{\C*\L /(\TmL+2*\L)} % TDMA rate

    % Slope params low prio
    \pgfmathsetmacro\Cql{.24*\C} % q' arriving rate
    \pgfmathsetmacro\bql{.3*\h} % q' burst
    \pgfmathsetmacro\Rql{\Rq} % q' TDMA rate

    % Bounding box for STA
    \draw[gray] ($(queue)-(3*\qw,-2.5*\qh)$)
        rectangle ++($(6+3*\qw,-9.5*\qh)$)
        node[anchor=south east,align=center,
        rotate=270]
        {802.11 STA\\with 802.1Qbv};
    
    % Draw opening gates
    \draw (0,\qy) -- (\deltaA, \qy)
        node[circle,fill=black,inner sep=0,
        minimum size=.25em] {};

    % Draw opening gates queue 2
    \draw (0,\qly) -- (\deltaA, \qly)
        node[circle,fill=black,inner sep=0,
        minimum size=.25em] {};

    \foreach \i in {0,...,\steps}{
        % Opening gates
        \draw  ($(\deltaA+\i*\TmL+\i*\L   ,\qy)$)
            -- ($(\deltaA+\i*\TmL+\i*\L+\L,\qy+.2)$);
        \draw  ($(\deltaA+\i*\TmL+\i*\L+\L   ,\qy)$)
        node[circle,fill=black,inner sep=0,
            minimum size=.25em] {}
            -- ($(\deltaA+\i*\TmL+\i*\L+\L+\TmL,\qy)$);
        \ifthenelse{\i=2}{}{%
            \node[circle,fill=black,inner sep=0,
                minimum size=.25em]
                at ($(\deltaA+\i*\TmL+\i*\L+\L+\TmL,\qy)$)
                {};
        }

        % Opening gates low prio. queue
        \draw  ($(\deltaA+\i*\TmL+\i*\L   ,\qly)$)
            -- ($(\deltaA+\i*\TmL+\i*\L+\L,\qly+.2)$);
        \draw  ($(\deltaA+\i*\TmL+\i*\L+\L   ,\qly)$)
        node[circle,fill=black,inner sep=0,
            minimum size=.25em] {}
            -- ($(\deltaA+\i*\TmL+\i*\L+\L+\TmL,\qly)$);
        \ifthenelse{\i=2}{}{%
            \node[circle,fill=black,inner sep=0,
                minimum size=.25em]
                at ($(\deltaA+\i*\TmL+\i*\L+\L+\TmL,\qly)$)
                {};
        }

        % TDMA increase
        \draw[color=gray,very thick]
            ($(\deltaA+\i*\TmL+\i*\L,\i*\h)$)
            --
            ($(\deltaA+\i*\TmL+\i*\L+\L,\i*\h+\h)$);
        
        \ifthenelse{\i=2}{%
            \node[anchor=south east,
                color=gray] at
                ($(\deltaA+\i*\TmL+\i*\L+.5*\L,\i*\h+.5*\h)$)
                {$\beta_{L_q,T_q}([t-\delta_q]^+)$};
        }{}

        \draw[color=gray,ultra thick]
            ($(\deltaA+\i*\TmL+\i*\L+\L,\i*\h+\h)$)
            --
            ($(\deltaA+\i*\TmL+\i*\L+\L+\TmL,\i*\h+\h)$);

    }

    % Lower envelope
    \draw[gray,thick,dashed] ($(\deltaA,0)$)
        --
        ($(\deltaA+\steps*\L+\L+\steps*\TmL+\TmL,\steps*\h+\h)$);
        % node[midway,anchor=north west]
        % {$\beta_q(t)$};

    % Lower envelope impacted
    \draw[color=DodgerBlue4,thick] ($(\deltaA+\llmax/\Rq,0)$)
        --
        ($(\deltaA+\steps*\L+\L+\steps*\TmL+\TmL,\steps*\h+\h-\llmax)$)
        node[pos=.8,anchor=north]
        {$\beta_q^*(t)$};

    % Low prio packet impact
    \draw [|-|,color=Firebrick4]
        ($(.15+ \deltaA+\steps*\L+\L+\steps*\TmL+\TmL,\steps*\h+\h-\llmax)$)
        --
        ($(.15+ \deltaA+\steps*\L+\L+\steps*\TmL+\TmL,\steps*\h+\h)$)
        node[midway,anchor=west] {$l_{\max}^{q'}$};

    % Lower envelope impacted for q'
    \pgfmathsetmacro\delayl{(\Rql*\deltaA+\bq)/(\Rql-\Cq)}
    \pgfmathsetmacro\ratel{(\Rql-\Cq)}
    \draw[color=Firebrick4,thick]
        ($(\delayl,0)$)
        --
        ($(5.5,\ratel*5.5 - \ratel*\delayl)$)
        node[pos=1,anchor=west]
        {$\beta_{q'}^*(t)$};

    % AXIS
    \draw[->] (0,0) -- (5.5,0)
        node[anchor=west] {$t$};
    \draw[->] (0,0) -- (0,\h*3)
        node[pos=.8,rotate=90,anchor=south] {bits};

    \draw[color=gray,very thick]
        (0,0) -- (\deltaA,0);

    % Specify the offset
    \draw[|-|,gray] ($(\qx+\qw+.1, \qly-.2)$)
        -- node[midway,anchor=north] {$\delta_{q'}$}
        ($(\qx+\qw+\deltaA-.1, \qly-.2)$);

    % Specify the gate opening time L
    \draw[|-|,gray] ($(\qx+\qw+\deltaA+.05, \qly-.2)$)
        -- node[midway,anchor=north] {$L_{q'}$}
        ($(\qx+\qw+\deltaA+\L-.05, \qly-.2)$);

    % Specify the gate period
    \draw[|-|,gray] ($(\qx+\qw+\deltaA+\TmL+\L, \qly-.2)$)
        -- node[midway,anchor=north] {$T_{q'}$}
        ($(\qx+\qw+\deltaA+2*\L+2*\TmL-.1, \qly-.2)$);

    % Specify q' envelope delay
    \pgfmathsetmacro\latencyl{%
        (\Rql*\deltaA+\bq)/(\Rql-\Cq)}
    \draw[|-|,color=Firebrick4]
        (0,-.25) -- (\latencyl,-.25)
        node[midway,anchor=north]
        {$\frac{R_{q'}\delta_{q'}+b_q}{R_{q'}-C_q}$};

    %%%%%%%%%%%%%%%
    % TWT windows %
    %%%%%%%%%%%%%%%

    % low priority queue q2
    \pgfmathsetmacro\twty{-5}
    \pgfmathsetmacro\twth{1}

    % TWT session time-axis
    \draw[->] (0,\twty) --
        ($(0,\twty)+(5.5,0)$)
        node[anchor=west] {$t$};
    \draw[->] (0,\twty) --
        ($(0,\twty)+(0,\twth)$)
        node[pos=.8,anchor=south,rotate=90] {freq.};

    % TWT sessions
    \foreach \i in {0,...,\steps}{
        \node (twtstart) at ($(\deltaA+\i*\TmL+\i*\L   ,\twty)$) {};

        % MU Tx box
        \draw[draw=black] (twtstart) {} rectangle ++(\L,\twth) ;
        % TWT session
        \draw[draw=gray,dotted]
            ($(twtstart)-(.1,0)$) {} rectangle ++($(\L,\twth)+(.2,.1)$) ;

        % First session
        \ifthenelse{\i=0}{%

            \draw[|-|] ($(twtstart)+(\L+.2,.05)$)
                -- ($(twtstart)+(\L+.2,\twth)$)
                node[anchor=south,pos=.5,
                rotate=270]
                {MU Tx};

            % packets
            \draw[fill=DodgerBlue1!40]
                ($(\deltaA+\i*\TmL+\i*\L,\twty+\twth/3)$) rectangle ++(\L/3,\twth/3);
            \draw[fill=Firebrick1!40]
                ($(\deltaA+\i*\TmL+\i*\L+\L/3,\twty+\twth/3)$) rectangle ++(\L/3,\twth/3);
            \draw[fill=DodgerBlue1!40]
                ($(\deltaA+\i*\TmL+\i*\L+2*\L/3,\twty+\twth/3)$) rectangle ++(\L/3,\twth/3);

        }{}

        % Second session
        \ifthenelse{\i=1}{%
            \draw[fill=DodgerBlue1!40]
                ($(\deltaA+\i*\TmL+\i*\L,\twty+\twth/3)$) rectangle ++(\L,\twth/3);

            % ALLOCATED RU
            \draw[|-|] ($(twtstart)+(\L+.2,\twth/3)$)
                -- ($(twtstart)+(\L+.2,2*\twth/3)$)
                node[anchor=west,pos=.5]
                {RU};
        }{}

        % Third session
        \ifthenelse{\i=2}{%

            \draw[fill=DodgerBlue1!40]
                ($(\deltaA+\i*\TmL+\i*\L,\twty+\twth/3)$) rectangle ++(2*\L/3,\twth/3);
            \draw[fill=Firebrick1!40]
                ($(\deltaA+\i*\TmL+\i*\L+2*\L/3,\twty+\twth/3)$) rectangle ++(\L/3,\twth/3);

            \node[anchor=west,align=left,
                color=gray]
                at ($(twtstart)+(\L+.1,\twth-.3)$)
                {\gls{rtwt}\\session};

            % MU Tx box
            \draw[black,dashed]
                ($(\deltaA+\i*\TmL+\i*\L,\twty+\twth)$) --
                ($(\deltaA+\i*\TmL+\i*\L,\qy)$);
            \draw[black,dashed]
                ($(\deltaA+\i*\TmL+\i*\L+\L,\twty+\twth)$) --
                ($(\deltaA+\i*\TmL+\i*\L+\L,\qy)$);

        }{}

    }

    %% Illustrate the allignment

\end{tikzpicture}}
    \vspace{-0.7em}
    \caption{An \protect\gls{sta} (middle) with
        two 802.1Qbv queues of high $q$ and low
        priority $q'$. The opening times
        of 802.1Qbv gates are aligned 
        with the \protect\gls{rtwt} sessions
        (bottom). The \protect\gls{sta} has
        a single RU in the MU Tx
        scheduled within each \protect\gls{twt}
        session. On top, 
        the resulting network calculus
        service curves $\beta_q^*,\beta_{q'}^*$.
    }
    \Description{The opening time of two queue gates aligned with multi-user transmissions inside TWT sessions.}
    \label{fig:concurrent}
    \vspace{-6mm}
\end{figure}

%\section{A Network Calculus approach}
%In this section we model the 802.1Qbv opening gates using Network Calculus~\cite{netcal8021qbv,leboudec}, and the wireless errors using stochastic scaling~\cite{stochastic-scaling}.

\section{Modelling 802.1Qbv as TDMA curves}
\label{sec:Qbv_tdma}
The \gls{tas} in 802.1Qbv decides when to open and close each queue using a \gls{gcl}. The latter defines both the periodicity $T_q$ of each queue and how long, $L_q$, it is open for transmission. Using \gls{tdma}, a user queue $q$ has a transmit service curve (cummulative bits processed over time), $\beta_{L,T}(t)$, given by \eqref{eq:tdma-service-curve}
\begin{equation}
    \beta_{L,T}(t)=
    C\cdot\max\left\{ \left\lfloor \frac{t}{T}\right\rfloor L,
    t-\left\lceil\frac{t}{T} \right\rceil (T-L) \right\}
    \label{eq:tdma-service-curve}
\end{equation}
with $C$ being the transmission rate and $T,L$ the periodicity and length of the \gls{tdma} transmission opportunities. The service curve in 802.1Qbv for a queue $q$ is given by $\beta_{L_q,T_q}([t-\delta_q]^+)$ where $\delta_q>0$ is the time at which queue $q$ gate is first opened.
%\begin{figure}[t]
%    \input{img/tdma.tex}
%    \caption{\gls{tdma} service curve 
%    $\beta_{L_q,T_q}([t-\delta_q]^+)$ for
%    a single 802.1Qbv queue $q$. The arriving
%    flow $A_q(t)$ waits an offset $\delta_q$
%    until the \gls{gcl} opens the gate for
%    the first time. Afterwards, the gates
%    open with $T_q$ period during slots of
%    length $L_q$. In dashed line we illustrate
%    the lower envelope of the \gls{tdma}
%    service curve.}
%\end{figure}

Note that in 802.11Qbv, two or more queues may have their gates opened concurrently\footnote{\ca{In this paper}, we consider 2 queues for brevity. Our analysis can be easily extended to $N$ concurrent queues.}. We use an affine lower envelope of the \gls{tdma} service curve~\eqref{eq:tdma-service-curve} to handle gate concurrency as shown in the  (top) dashed line of Figure~\ref{fig:concurrent}. The \gls{tdma} affine lower envelope for queue $q$ in 802.1Qbv is
the rate-latency function
\begin{equation}
    \beta_q(t)=\frac{C\cdot L_q}{T_q+L_q}[t-\delta_q]^+.
\end{equation}

\section{Concurrent 802.1Qbv opened gates}
\label{subsec:concurrent}

When high priority queue $q$ and low priority queue $q'$ are opened concurrently, packets of $q$ are transmitted ahead of those from $q'$.

%We can tell $q$ incoming traffic is governed by a strict arrival curve $\gamma_{C_q,b_q}(t)=C_q\cdot t + b_q$, and queue $q'$ by a strict arrival curve $\gamma_{C_{q'},b_{q'}}(t)$. Therefore, it is possible to know how the service curve of the low priority queue $q'$ is impacted by the high priority traffic.

\begin{corollary}[802.1Qbv High priority service curve]
    \label{cor:HPservice}
    Given two queues $q,q'$ whose gates
    are opened concurrently; the high
    priority queue $q$ has a strict service
    curve
    \begin{equation}
        \label{eq:HPservice}
        \beta_q^*(t) = R_q\left[t-\left(\delta_q+\frac{l_{\max}^{q'}}{R_q}\right) \right]^+
    \end{equation}
    with $R_q=\tfrac{C L_q}{T_q+L_q}$ and
    $l_{\max}^{q'}$ the maximum packet size
    of the low priority queue $q'$.
\end{corollary}
\begin{proof}
    We mimic the proof of
    \cite{leboudec}[Proposition 1.3.4].
    Take the affine lower envelope of queue
    $q$ defined
    in~\eqref{eq:tdma-service-curve}. Consider
    $s$ as the start of the backlog period of
    queue $q$, hence, the arrival and departing
    curve at queue $q$ is the same at that time
    $D_q(s)=A_q(s)$.

    For 802.1Qbv is non-preemptive, if a
    low priority packet from $q'$ arrives before
    a high priority packet from $q$ in the
    interval $(s,t]$; the high priority packet
    will wait, i.e.
    \begin{equation}
        D_q(t)-D_q(s)\geq \beta_q(t-s)-l_{\max}^{q'}
    \end{equation}
    given $D_q(s)=A_q(s)$ and
    $A,D\in\mathcal{F}=\{f: \mathbb{R}^+\to \mathbb{R}, \forall t\geq s:\ f(t)\geq f(s), f(0)=0\}$;
    we know
    \begin{equation}
        D_q(t)\geq A_q(s)+
        \left[ R_q\left[ (t-s)-\delta_q \right]^+  - l_{\max}^{q'} \right]^+
        \label{eq:depart-high}
    \end{equation}
    If $(t-s)>\delta_q$
    \eqref{eq:depart-high} becomes
    \begin{equation}
        D_q(t)\geq
            A_q(s)+
            R_q\left[ (t-s)- \left(\delta_q+ \frac{l_{\max}^{q'}}{R_q} \right) \right]^+
    \end{equation}
    If $(t-s)\leq\delta_q$ 
    \eqref{eq:depart-high} becomes
    $D_q(t)\geq A_q(s)$.
    That is, the 
    departing flow of the high priority
    queue $q$ satisfies
    \begin{equation}
        D_q(t)\geq A_q\otimes \beta_q^*(t)
    \end{equation}
    with $\beta_q^*(t)$ being the
    rate-latency service curve with
    rate $R_q$ and latency 
    $\delta_q+\frac{l_{\max}^{q'}}{R_q}$.
\end{proof}

Similarly, it is also possible to obtain the
service curve for the traffic in the
low priority queue $q'$ when such queue is
opened concurrently with the higher priority
queue $q$.

\begin{corollary}[802.1Qbv Low priority service curve]
    \label{cor:LPservice}
    Given two queues $q,q'$ whose gates
    are opened concurrently;
    if the flow at $q$ has
    a strict affine arrival curve
    $\gamma_{C_q,b_q}$, then the low
    priority queue $q'$ has a strict service
    curve
    \begin{equation}
        \label{eq:LPservice}
        \beta_{q'}^*(t)=
        \left(R_{q'}-C_q \right)
        \left[
            t
            - \frac{ R_{q'}\delta_{q'}+ b_q}{R_{q'}-C_q}
        \right]^+
    \end{equation}
    with $R_{q'}=\tfrac{C L_{q'}}{T_{q'}+L_{q'}}$.
\end{corollary}
\begin{proof}
    We again mimic the proof presented in \cite{leboudec} [Proposition 1.3.4].
    Take the affine lower envelope of
    queue $q'$ as defined
    in~\eqref{eq:tdma-service-curve}. Consider
    $s'$ the start of the busy period, that is,
    $s'<s$ with $s$ the start of the backlog
    period. In the interval $(s',t]$ the
    low priority queue departing traffic
    satisfies
    \begin{equation}
        D_{q'}(t)-D_{q'}(s')=
        \beta_{q'}(t-s') -
        \left[ D_q(t) - D_q(s') \right]
        \label{eq:departing-low}
    \end{equation}
    Given that at $s'$ there is no backlog,
    and the strict arrival curve for the
    high priority queue $q$, we know
    \begin{multline}
        D_q(t)-D_q(s')=D_q(t)-A_q(s')\\
        \leq A_q(t)-A_q(s')
        \leq \gamma_{C_q,b_q}(t-s')
        \label{eq:low-depart-difference_prior}
    \end{multline}
    Then, we substitute ~\eqref{eq:low-depart-difference_prior}
    in~\eqref{eq:departing-low} to obtain
    \begin{multline}
        D_{q'}(t)-D_{q'}(s')
        = D_{q'}(t)-A_{q'}(s')\\
        \geq \left[ R_{q'}\left[
            (t-s)-\delta_{q'}
        \right]^+
        - \left( C_q(t-s')+b_q \right)
        \right]^+
        \label{eq:low-depart-difference}
    \end{multline}
    Again, depending on $(t-s)$ the above
    expression changes. For
    $(t-s)>\delta_{q'}$,
    \eqref{eq:low-depart-difference}~becomes
    \begin{equation}
        D_{q'}(t)\geq A_{q'}(s')+\left[
            \left( R_{q'}-C_q \right)
            (t-s') - \left( R_{q'}\delta_{q'}
            + b_q\right)
        \right]^+
    \end{equation}
    and with $(t-s')\leq\delta_{q'}$,
    inequality~\eqref{eq:low-depart-difference}
    becomes $D_{q'}(t)\geq A_{q'}(s')$.
    Overall, we can tell that the departing
    flow at the low priority queue $q'$
    satisfies
    \begin{equation}
        D_{q'}(t)\geq A_{q'}\otimes\beta_{q'}^*(t)
    \end{equation}
    with $\beta_{q'}^*(t)$ being the
    rate-latency service curve with rate
    $R_{q'}-C_q$ and latency
    $\tfrac{R_{q'}\delta_{q'}+b_q}{R_{q'}-C_q}$.
\end{proof}

\section{Retransmissions in 802.11}
\label{sec:losses}

So far, we have obtained the service curves $\beta_q^*(t),\beta_{q'}^*(t)$ of high/low priority traffic upon the opened/closed 802.1Qbv gates. By aligning gate opening with the \gls{rtwt} sessions as depicted in Figure~\ref{fig:concurrent}, channel contention is avoided since the \gls{sta} has a dedicated \gls{ru} in 802.11ax \gls{mu} Tx, and the \gls{rtwt} session access is restricted.

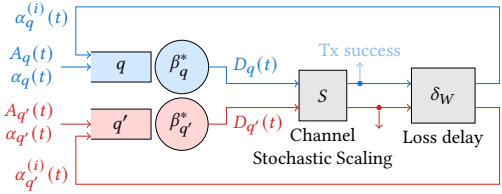
\begin{figure}[t]
    \centering
    \scalebox{0.8}{\begin{tikzpicture}

    % queue
    \pgfmathsetmacro\qy{-1.7}
    \pgfmathsetmacro\qx{-.5}
    \pgfmathsetmacro\qw{.5} % 1/2 queue width
    \pgfmathsetmacro\qh{.25} % 1/2 queue height
    \coordinate (queue) at (\qx,\qy);
    \path[fill=DodgerBlue1!20,draw] ($(queue)+(-\qw,\qh)$) --
        ($(queue)+(\qw, \qh)$) --
        node (betaqball) {}
        ($(queue)+(\qw,-\qh)$) --
        ($(queue)+(-\qw,-\qh)$);
    \node[circle,draw,fill=DodgerBlue1!20]
        (ballq)
        at ($(betaqball.east)+(.4,0)$)
        {$\beta_{q\phantom{'}}^*$};
    \node at (queue) {$q$};

    % Delay box
    %\node[rectangle,draw]

    % Arrival flow
    \draw[->,color=DodgerBlue3]
        ($(queue)-(2*\qw,0)$)
        node[anchor=east,align=center]
        {$A_q(t)$\\$\alpha_q(t)$}
        -- ($(queue)-(\qw,0)$);

    % low priority queue q2
    \pgfmathsetmacro\qly{-2.65}
    \pgfmathsetmacro\qlx{-.5}
    \pgfmathsetmacro\qlw{.5} % 1/2 queue width
    \pgfmathsetmacro\qlh{.25} % 1/2 queue height
    \pgfmathsetmacro\llmax{.2} % max low packet
    \coordinate (queuel) at (\qlx,\qly);
    \path[fill=Firebrick1!20,draw] ($(queuel)+(-\qw,\qh)$) --
        ($(queuel)+(\qlw, \qlh)$) --
        node (betaqlball) {}
        ($(queuel)+(\qlw,-\qlh)$) --
        ($(queuel)+(-\qlw,-\qlh)$);
    \node[circle,draw,fill=Firebrick1!20]
        (ballql)
        at ($(betaqlball.east)+(.4,0)$)
        {$\beta_{q'}^*$};
    \node at (queuel) {$q'$};

    % Scaling box
    \node (middle) at ($(0,.5*\qy+.5*\qly)$) {};
    \node (rightballql) at
        ($(ballql.east)+(1.9,0)$) {};
    \node[rectangle,draw,fill=gray!20,
        inner sep=1em] (scaling) at
        (rightballql |- middle)
        {$S$};
    \node[anchor=north,align=center]
        at (scaling.south)
        {Channel\\Stochastic Scaling};

    % Delay box
    \node[rectangle,draw,fill=gray!20,
        inner sep=1em]
        (delay) at ($(scaling)+(2,0)$)
        {$\delta_W$};
    \node[anchor=north] at (delay.south) {Loss delay};

    % Arrival flow low priority
    \draw[->,color=Firebrick3]
        ($(queuel)-(2*\qlw,0)$)
        node[anchor=east,align=center]
        {$A_{q'}(t)$\\$\alpha_{q'}(t)$}
        -- ($(queuel)-(\qlw,0)$);

    %%%%%%%%%%%%%%%%%%
    %%%% ARROWS %%%%%%
    %%%%%%%%%%%%%%%%%%

    \draw[->,color=DodgerBlue3] (ballq) -|
        ($(scaling.west)+(-1.2,.2)$) --
        node[pos=0,anchor=south west]
        {$D_q(t)$}
        ($(scaling.west)+(0,.2)$);

    \draw[->,color=Firebrick3] (ballql) -|
        ($(scaling.west)-(1.2,.2)$) --
        node[pos=0,anchor=north west]
        {$D_{q'}(t)$}
        ($(scaling.west)-(0,.2)$);

    \draw[->,color=DodgerBlue3]
        ($(scaling.east)+(0,.2)$)
        -- ($(delay.west)+(0,.2)$)
        node[pos=0.2,circle,fill=DodgerBlue3,
        minimum size=.1em,inner sep=.1em]
        (exit) {};
    \draw[->,color=DodgerBlue3!50]
        (exit) -- ($(exit)+(0,.35)$)
        node[anchor=south]
        {Tx success};

    \draw[->,color=Firebrick3]
        ($(scaling.east)-(0,.2)$)
        -- ($(delay.west)-(0,.2)$)
        node[pos=0.5,circle,fill=Firebrick3,
        minimum size=.1em,inner sep=.1em]
        (exitl) {};
    \draw[->,color=Firebrick3]
        (exitl) -- ($(exitl)-(0,.35)$);

    \draw[->,color=Firebrick3]
        ($(delay.east)-(0,.2)$)
        -- ($(delay.east)+(.4,-.2)$)
        |- ($(queuel.west)-(1.5*\qlw,4*\qlh)$)
        |- ($(queuel.west)-(\qlw,.2)$)
        node[pos=.1,anchor=east]
        {$\alpha_{q'}^{(i)}(t)$};

    \draw[->,color=DodgerBlue3]
        ($(delay.east)+(0,.2)$)
        -- ($(delay.east)+(.4,.2)$)
        |- ($(queue.west)+(-1.5*\qlw,4*\qlh)$)
        |- ($(queue.west)+(-\qlw,.2)$)
        node[pos=.1,anchor=east]
        {$\alpha_{q}^{(i)}(t)$};

\end{tikzpicture}}
    \vspace{-5mm}
    \caption{When the $q,q'$ are opened, packets 
    $D(t)$ traverse the stochastic wireless
    channel $S$. On Tx error, %it takes $\delta_W$ to realize the packet loss. 
    $\delta_W$ represents the time it takes to realize the packet loss. 
    Then, packets are fed-back for retransmission leading to re-entrant arrivals $\alpha^{(i)}(t)$.}
     \Description{Two queues exiting the same channel, hence suffering from the same stochastic scaling and loss delay, causing that the failed traffic needs to be transmitted again.}
    \label{fig:scaling}%
    \vspace{-5mm}
\end{figure}

However, packet errors may still occur due to channel impairments resulting in retransmissions. To deal with the latter, we use the stochastic scaling approach proposed in~\cite{stochastic-scaling}. The idea is to send the departing traffic of 802.1Qbv $D_q(t),D_{q'}(t)$ to a wireless channel modeled with a scaling process $S$ as shown in Figure~\ref{fig:scaling}. If the Tx does not succeed, the \gls{sta} waits time $W$ (as maximum) before retransmission of the i\textsuperscript{th} packet. This process is modelled with the scaling curve $\delta_W$.

Following~\cite{stochastic-scaling}, we consider a binary symmetric channel modelled with stochastic scaling process $S(b)=\sum_i^b X_i,\ X_i\overset{\mathrm{iid}}{\sim} \text{Bernoulli}(p)$, with $p$ the
loss probability.
We now define a stochastic scaling
curve $S^\varepsilon(b)=pb+1-\varepsilon$
that, according to~\cite[Theorem~1]{stochastic-scaling},
satisfies
\begin{equation}
    \mathbb{P}\left( \sup_{0\leq a\leq b} \left\{ S(b)-S(a)-S^\varepsilon(b-a) \right\}\leq0\right)\geq 1-\varepsilon,\quad \forall b\geq0
\end{equation}
that is, the queue experiencing Tx errors $S(D(t))$ is upper bounded by the outgoing traffic (which is bounded by $S^\varepsilon(D(t))$) with probability $1-\varepsilon$.

The scaled flow $S(D(t))$ will result in
retransmissions, namely, we have
$\alpha^{(i)}(t),\ i=0,1,\ldots,N$ being the
arrival curves for the 
i\textsuperscript{th} retransmissions{\color{black}, where $\alpha^{(0)}=A(t)=\gamma_{r,b}$ represents the original input flow}. The higher the Rtx index, the higher the priority, thus each retransmission flow has arrival and service curves
\begin{equation}
    \beta_q^{(i)}=\left[
        \beta-\sum_{k=i+1}^N \alpha^{k}
    \right]^+,
    \qquad
    \alpha_q^{(i)}=
    S^\varepsilon\left(
        \alpha^{(i-1)}\oslash\beta^{(i-1)}
    \right)\oslash \delta_W.
\end{equation}

%{\color{red}TODO update from here and
%say that the above results in a system
%of equations, and refer to \cite{stochastic-scaling}. Then, mention that the theorem in
%the paper ensures (1-epsion)N reliability,
%and show that the delay is obtained using
%h(...).}

%We can then define a stochastic scaling curve for $S(b)$ to know how likely the packets are retransmitted. 
{\color{black}
%As \cite[Section~IV.C]{stochastic-scaling} demonstrates, above definitions result in a system of equations for obtaining the arrival curves of all retransmissions. 

Once the system is solved, we define the aggregated arrival curve for the queue $q$ as 
%\begin{equation*}
    $\alpha^{(Tot)}_q=\sum_{j=0}^N \alpha^{(j)}_q.$
%\end{equation*}
At this point, given the aggregated arrival curve and the service curve, Network Calculus provides bounds to obtain the maximum delay experienced by queue $q$ using the following expression:
\begin{equation*}
    h(\alpha,\beta)\, \raisebox{.5pt}{:}\!= \sup_{s\geq0} \left\{ \inf\{u\geq0: \alpha(s)\leq\beta(s+u)\} \rule{0pt}{1em} \right\}.
\end{equation*}

Note that previous curves are derived from the stochastic scaling process that model the channel, hence, they are not deterministic. In fact, as \cite{stochastic-scaling} highlights, the reliability of the delay bound (i.e., the probability that it holds) will be 
\begin{equation}
    \mathbb P\left( d(t) \leq h(\alpha^{(Tot)},\beta) \right)
    \geq (1-\varepsilon)^N \qquad \forall t\in\mathbb R^+,
\end{equation}
with $d(t)$ representing the real delay experienced by packets at time $t$.
}

\section{Results}
\label{sec:results}
\begin{figure}[t]
    \scalebox{0.8}{\input{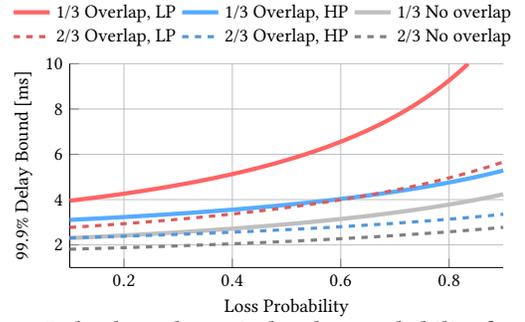}}
    \vspace{-5mm}
    \caption{
    Delay bound vs.
    wireless loss probability
    for different $\tfrac{L}{T}$ ratios.
    The Delay Bound holds
    with a 99.9\%
    probability.
    We test non overlapping
    setups with single 
    802.1Qbv queues (gray),
    and overlapping setups
    with concurrent high (HP)
    and low priority (LP) queues
    (blue and red).
    \Description{Curves representing the delay bound in function of the loss probability; main results are detailed in the text.}
}
    \label{fig:results}
    \vspace{-5mm}
\end{figure}

To highlight the importance of these results, we analyze two scenarios: with $(i)$ $T=3$, $L=1$ and $(ii)$  $T=3$, $L=2$.
In both, we calculate the aggregated arrival curves for one non-overlapping queue and two overlapping queues (one with higher priority than the other).
The rest of the parameters are $C=10,\ \alpha^{(0)}=\gamma_{0.1,0.001}$ and $\ N=3$. Additionally, the violation probability of the stochastic scaling curve $\varepsilon$ is $3.3344\cdot10^{-4}$.

Figure~\ref{fig:results} shows the maximum delay experienced by the queues as a function of the loss probability $p$. Note that these bounds hold with a $0.999$ probability. We observe that the delay is reduced with greater overlap. The high priority queue is less impacted compared to the low priority queue. Moreover, comparing the $L/T=1/3$ non-overlapping case with $2/3$ overlapping case, the high priority queue in the latter case experiences less delay at the expense of the low priority queue.

\section{Conclusions and future work}
\label{sec:conclusions}

We have presented stochastic bounds for \gls{tsn} packets in a 802.11 deployment, with queue concurrency and retransmissions due to channel impairments. Our expressions assume the alignment of 802.1Qbv queues with 802.11 \gls{twt} sessions.

This paper provides a preliminary, simplified assessment of dense Industry 4.0 \gls{tsn} wireless deployments. We intend to extend our analysis to more complex scenarios and develop a proof of concept experiments to validate our results.

\section*{Acknowledgment}
This work has been partially funded by the European Commission Horizon Europe SNS JU PREDICT-6G (GA 101095890) Project and the Spanish Ministry of Economic Affairs and Digital Transformation and the European Union-NextGenerationEU through the UNICO 5G I+D 6G-EDGEDT and 6G-DATADRIVEN.
\bibliographystyle{acm}%{ACM-Reference-Format}
\bibliography{bibliography.bib}


\end{document}